\documentclass[pra,showpacs,twocolumn]{revtex4}
\usepackage{dcolumn}
\usepackage{bm}
\usepackage{amssymb}
\usepackage{amsmath}
\usepackage{amsfonts}
\usepackage[]{graphicx}
\begin{document}
\bibliographystyle{prsty}
\title{Plasmonic interferometry: probing launching dipoles in scanning-probe plasmonics}
\author{Oriane Mollet$^1$, Guillaume Bachelier$^1$, Cyriaque Genet$^2$, Serge Huant$^1$, Aur\'{e}lien Drezet$^{1\dag}$}
\address{(1)Institut N\'{e}el, CNRS and Universit\'{e} Joseph Fourier,
BP166, 38042 Grenoble Cedex, France}
\address{(2)ISIS, UMR 7006, CNRS-Universit\'e de Strasbourg, 8, all\'ee Monge, 67000 Strasbourg, France}
\begin{abstract}
We develop a semi-analytical method for analyzing surface plasmon
interferometry using near-field scanning optical sources. We compare
our approach to Young double hole interferometry experiments using
scanning tunneling microscope (STM) discussed in the literature and
realize experiments with an aperture near-field scanning optical
microscope (NSOM) source positioned near a ring like aperture slit
milled in a thick gold film.   In both cases the agreement between
experiments and model is very good. We emphasize the role of dipole
orientations and discuss the role of magnetic versus electric dipole
contributions to the imaging process as well as the directionality
of the effective dipoles associated with the various optical and
plasmonic sources.
\end{abstract}

\pacs{} \maketitle
\section{Introduction}
\indent Surface plasmons (SPs)~\cite{BarnesNature2003,emrs} and
NSOM~\cite{Novotny} share a long common historical background which
goes back to the birth of both research fields. Indeed, SPs,
collective electronic excitations bounded at a metal dielectric
interface, belong to the family of electromagnetic waves which are
evanescent in the direction normal to the interface. As such, a near
field probe located in the vicinity of the metal can be used to
either record or excite SP waves. Here we will focus on the
application of such a scanning probe to plasmonic
interferometry using holes and slits.\\
\indent First, it is worth reminding that photon STM
(PSTM)~\cite{Novotny} was used long ago to probe SP waves externally
excited with a tapered fiber tip located in the near field of the
plasmon~[4-7]. Recently, the method was applied to observe SP
interference fringes in Fabry-Perot resonators such as
nanowires~\cite{Ditlbacher} and plasmonic corrals~[9-11]. The
reciprocal regime i.e., the excitation of SP waves with a scanned
nano-antenna, is less common but was also reported using NSOM with
aperture optical tips~\cite{Novotny} by Hecht et
al.~\cite{Hecht,2b}. The method was soon applied to observe SP
interference fringes inside lithographed cavities \cite{3c,3b} or
between pairs of holes in metal film~\cite{Sonnichsen}. This was
also successfully used to launch SPs to excite the luminescence of
quantum dots at low temperature~\cite{brun}, thereby entering into
the realm of quantum plasmonics. In this context, a new family of
active NSOM tips using a single quantum emitter (such as a single
Nitrogen vacancy center in a nano-diamond) glued at the apex of an
optical tip~\cite{12,13}, was recently used to generate quantum SP
states propagating along a
metal film or inside plasmonic corrals [20-22].\\
\indent In all these studies the dipolar approximation for
describing the electromagnetic field generated by the NSOM tip has
been applied with success in agreement with previous optical
characterization measurements and models of aperture
tips~[19,23-26]. SPs launched from such aperture NSOM tips show
characteristics dipolar profiles [12,14-17,20,21]. This is
reminiscent of the seminal Sommerfeld antenna theory introducing
surface waves~\cite{SommerfeldAP1909}, which was recently extended
to thin metal films supporting SP waves launched by point-like
radiating dipoles~\cite{2b,NikitinPRL2010,Genet}. Experimental
demonstration in both the near field or far field using the
so-called leakage radiation microscopy (LRM) with aperture or active
NSOM tips~\cite{2b,15,19,Genet,17} confirms these findings, in
particular the fact that a NSOM tip behaves essentially as an
in-plane
radiating dipole.\\
\indent Interestingly, similar features were obtained using STM
tips~\cite{STM1,STM2}. In these experiments, inelastic electrons
tunneling through a tip/metal film junction generate
photoemission~\cite{STM3,STM4} which, in turn, couples into SPs
propagating along the metal/air interface. Modeling of such a system
involves a vertical transition dipole aligned with the STM tip
revolution axis~\cite{marty,STM1} in agreement with LRM
images~\cite{STM1,STM2}.\\
\indent The potentialities of both NSOM and STM scanning sources for
plasmonics have however not be fully appreciated despite the fact
that these sources provide a natural way for mapping  the localized
density of states (LDOS) of photonic  and plasmonic modes around
nanostructures~\cite{Gerard,3,24}. This strongly motivates the
present work that is devoted to a better understanding of the
interactions between scanning excitation sources and a plasmonic
environment. Moreover, due to the precise position control of NSOM
or STM tips, this SP excitation method provides a natural way to
study SP interferometry different from the PSTM
approach~[9-11,40,41]. An experiment using a STM tip to launch SP
waves in the direction of two subwavelength holes milled in a thick
gold film was recently reported~\cite{sample}. The holes, acting
together as a pair of coherent sources induced optical fringes in
the Fourier plane of a microscope objective. This is reminiscent of
previous adaptations of Young double slits experiments with noble
metals~\cite{Gan,Kuzmin} in which a SP contribution was clearly
involved in the far-field fringe visibility and phase shift.\\
\indent In the present work, we study SP interferometry both
theoretically and experimentally using an aperture NSOM launching
SPs inside a circular corral. We also discuss the recent experiments
realized with a STM tip and two subwavelength holes~\cite{sample}
and extend our discussion to larger hole diameters. We emphasize the
role of polarization and tip dipole orientations on SP
interferometry in both configurations. In particular, we show that
the method provides an elegant way to discuss experimentally the old
problem of the orientation of the aperture dipoles associated with
holes or slits in an opaque metal film~\cite{Rotenberg,Yin,Yi}.
This, we claim, could play an important role in the context of
polarization conversion involving an angular momentum exchange
between SP modes, photons, and nanostructuration (see for
example~[11,46-48]).
\section{Electric and magnetic dipoles as SP launchers}
\indent We start with the description of SPs launched by a NSOM or
STM tip along a thick metal film. SP modes propagating along a flat
interface $z=0$ are inhomogeneous transverse magnetic waves which
are completely defined in each medium $j=0,1$ (air and metal) by a
characteristic Debye function
$\Psi_{\textrm{SP};j}(x,y,z)e^{-i\omega t}$ obeying the usual
Helmholtz equation
$[\boldsymbol{\nabla}^2+k^2\varepsilon_j]\Psi_{\textrm{SP};j}=0$,
where $k=2\pi/\lambda=\omega/c$ is the wavevector in vacuum and
$\varepsilon_j(k)$ the dielectric permittivity of each medium. From
such a SP characteristic function, the electric displacement
$\mathbf{D}_{\textrm{SP};j}$ and magnetic (induction) field
$\mathbf{B}_{\textrm{SP};j}$  are defined (in the Heaviside system
of units) by:
\begin{eqnarray}
\mathbf{D}_{\textrm{SP};j}(x,y,z)=\boldsymbol{\nabla}\times\boldsymbol{\nabla}\times[\mathbf{\hat{z}}\Psi_{\textrm{SP};j}(x,y,z)]\nonumber\\
\mathbf{B}_{\textrm{SP};j}(x,y,z)=-ik\boldsymbol{\nabla}\times[\mathbf{\hat{z}}\Psi_{\textrm{SP};j}(x,y,z)],\label{a}
\end{eqnarray} $z$ defining the normal direction to the interface and $\mathbf{\hat{z}}$ oriented from the metal side ($z<0$) to the air side ($z>0$). Using boundary conditions at the interface and infinity we
write $\Psi_{\textrm{SP};j}(x,y,z)=\Phi_{\textrm{SP}}(x,y,)f_j(z)$
where $f_0(z)=e^{ik_0 z}$ and $f_1(z)=e^{-ik_1 z}$ define
exponential decays in both media justifying the SP confinement at
the interface. Also $\Phi_{\textrm{SP}}(x,y,)$ fulfills the
bidimensional Helmoltz equation
$[\boldsymbol{\nabla}_{||}^2+k^2_{\textrm{SP}}]\Phi_{\textrm{SP}}(x,y)=0$
 , with $\boldsymbol{\nabla}_{||}=\frac{\partial}{\partial x}\mathbf{\hat{x}}+\frac{\partial}{\partial y}\mathbf{\hat{y}}$ characterized by the usual SP wave vector
$k_{\textrm{SP}}=k\sqrt{\frac{\varepsilon_0\varepsilon_1}{\varepsilon_0+\varepsilon_1}}=kn_{\textrm{SP}}+i/(2L_{\textrm{SP}})$
($n_{\textrm{SP}}$ and $L_{\textrm{SP}}$ are respectively the SP
effective index and propagation length) together with
$k_j=\sqrt{k^2\varepsilon_j-k^2_{\textrm{SP}}}$ (to obtain
attenuation we impose
$\textrm{Imag}[k_j]>0$)~\cite{BarnesNature2003}.\\
For the present purpose we consider the SP field generated by a
point-like dipole located near the interface. A rigorous calculation
involves an evaluation of the Sommerfeld integral obtained by
continuation in the complex plane~\cite{2b,SommerfeldAP1909,Genet},
but this goes beyond the purpose of the present paper. Here, we are
only interested in the so-called singular or polar contribution
which is associated with the propagating SP wave and dominates the
far-field~\cite{NikitinPRL2010,Genet}. From symmetry considerations
we can directly obtain the characteristic Debye potentials
$\Phi_{\textrm{SP}}$ (of argument
$\mathbf{r}=x\mathbf{\hat{x}}+y\mathbf{\hat{y}}$) for a vertical and
horizontal dipole. Using Hankel functions these potentials are
respectively:
\begin{eqnarray}
\Phi_{\textrm{SP},\bot}(\mathbf{r})=\eta_\bot p_z
H_0^{(+)}(k_{\textrm{SP}}|\mathbf{R}|)\nonumber\\
\Phi_{\textrm{SP},||}(\mathbf{r})=\eta_{||}
H_1^{(+)}(k_{\textrm{SP}}|\mathbf{R}|)\mathbf{\hat{R}}\cdot\mathbf{p}_{||},
\end{eqnarray}
with $p_z$ and $\mathbf{p}_{||}$ the vertical, respectively
horizontal, dipole components
 and $\mathbf{R}=\mathbf{r}-\mathbf{r}_0$ if the tip is located at
$\mathbf{r}_0=x_0\mathbf{\hat{x}}+y_0\mathbf{\hat{y}}$. The coupling
constants
$\eta_{\bot}=\frac{1}{2}\frac{\varepsilon_1^2}{\varepsilon_1^2-\varepsilon_0^2}e^{ik_0h}$
and $\eta_{||}=i\frac{k_0}{k_{\textrm{SP}}}\eta_{\bot}$ depend
explicitly on the dipole height $z=h$ over the metal surface and are
evaluated \begin{figure}[h]
\begin{center}
\includegraphics[width=8.5cm]{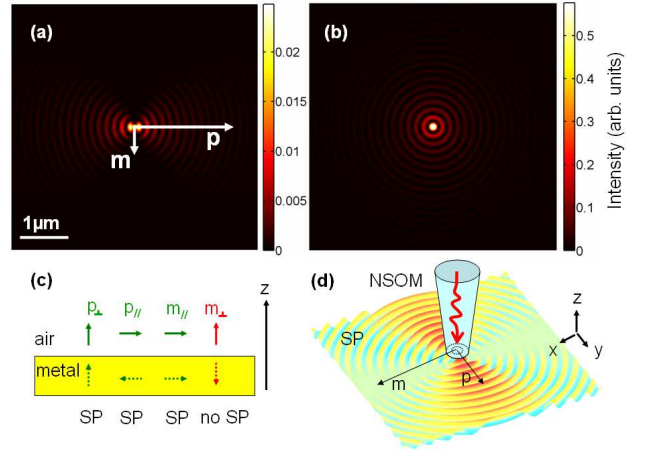}
\caption{ (a), respectively (b): SP intensity field, i.e.
$(\textrm{Re}[E_z])^2)$  generated by a  point-like  horizontal,
respectively vertical, dipolar source. In (a) the source can be
either an electric or magnetic in-plane dipole. The horizontal,
respectively vertical, white arrow represents the direction and
relative intensity of the equivalent electric, respectively
magnetic, dipoles. In (b) the intensity corresponds to a vertical
dipole of same amplitude as the electric dipole of (a). (c): table
summarizing the possibility for exciting SPs with dipoles located
near a metal film. The dipoles directions and those of their
(quasi-static) image dipoles in the thick film are depicted to guide
the reasoning. (d): representation of the SP field generated by an
aperture NSOM source located near a metal film. The direction of the
two effective electric and magnetic dipoles are represented by black
arrows. The intensity map corresponds to (a).       }
\label{figure5}
\end{center}
\end{figure}
using the residue theorem (see also~\cite{Genet} for
details).\\
\indent A few remarks are here relevant: first the coupling
constants decay exponentially as $h$ increases showing that one must
be very close from the surface in order to excite SP waves. Second,
the relation between $\eta_{\bot}$ and $\eta_{||}$ implies that it
is much easier to excite SPs with a vertical dipole than with an
in-plane dipole since the ratio $\frac{k_0}{k_{\textrm{SP}}}$ is in
general smaller than one. This agrees with a simple qualitative
argument comparing the image dipoles for both configurations.  For
example, at a wavelength $\lambda=647$ nm for an air/gold interface
we obtain $\frac{k_0}{k_{\textrm{SP}}}\simeq0.27$ using
$\varepsilon_0\simeq1$, $\varepsilon_1\simeq -13.6482 + 1.0352i$
\cite{Johnson}. The probability of SP excitation  for an horizontal
dipole amounts therefore to only
$(\frac{k_0}{k_{\textrm{SP}}})^2=7\%$ of the emission obtained for a
vertical dipole of the same amplitude. This is particularly relevant
for NV-based NSOM tips~[18-22], where the fixed transition dipoles
are oriented in a uncontrollable way in the diamond crystal. We also
point out that the above analysis is done by assuming the film
thickness $d$ large enough to neglect the effect of the second
interface~\cite{emrs}, typically $d\gtrsim70$ nm for gold film.
However, the main reasoning keeps its validity for thinner films
where leakage radiation is involved. In particular the ratio
formulas between $\eta_{||}$ and $\eta_{\bot}$ or $W_{||}$ and
$W_{\bot}$ are
kept unchanged even though the dispersion relation for $k_{\textrm{SP}}$ is modified~\cite{emrs}.\\
\indent Most importantly for the present work, we stress that the
asymptotic behavior of the singular Hankel functions
$H_0^{(+)}(k_{\textrm{SP}}|\mathbf{R}|)$ and
$H_1^{(+)}(k_{\textrm{SP}}|\mathbf{R}|)$ goes like
$\frac{e^{ik_{\textrm{SP}}|\mathbf{R}|}}{\sqrt{R}}$ if $R>>\lambda$.
Therefore, inserting these expressions in Eq.~1 we obtain the so
called Zenneck SP mode:
$\mathbf{D}_{\textrm{SP};j}(\mathbf{r},z)\simeq\Phi_{\textrm{SP}}(\mathbf{r})f_j(z)\mathbf{g}_j$,
$\mathbf{B}_{\textrm{SP};j}(\mathbf{r},z)\simeq\Phi_{\textrm{SP}}(\mathbf{r})f_j(z)k_{\textrm{SP}}k\mathbf{\hat{R}}\times\mathbf{\hat{z}}$
with
$\mathbf{g}_0=-k_{\textrm{SP}}k_0\mathbf{\hat{R}}+k_{\textrm{SP}}^2\mathbf{\hat{z}}$,
and
$\mathbf{g}_1=+k_{\textrm{SP}}k_1\mathbf{\hat{R}}+k_{\textrm{SP}}^2\mathbf{\hat{z}}$.
More generally, using such a Debye potential formalism, it is easy
to obtain from Eqs. 1,2 the plasmonic dyadic Green function defined
by
$\mathbf{D}_{\textrm{SP};j}(\mathbf{r},z)=\overline{\mathbf{G}}_{\textrm{SP}}(\mathbf{r},z,\mathbf{r}_0,h;j)\cdot\mathbf{p}$
and
\begin{eqnarray}
\overline{\mathbf{G}}_{\textrm{SP}}(\mathbf{r},z,\mathbf{r}_0,h;j)=
\eta_{\bot}f_j(z)\{-\delta_j i k_jk_{\textrm{SP}}
\nonumber\\ \cdot H_1^{(+)}(k_{\textrm{SP}}|\mathbf{R}|)\hat{\mathbf{R}}\otimes\hat{\mathbf{z}}+k_{\textrm{SP}}^2H_0^{(+)}(k_{\textrm{SP}}|\mathbf{R}|)\hat{\mathbf{z}}\otimes\hat{\mathbf{z}}\nonumber\\
+i\frac{k_0}{k_{\textrm{SP}}}[\delta_j i
k_jk_{\textrm{SP}}\frac{H_0^{(+)}(k_{\textrm{SP}}|\mathbf{R}|)-H_2^{(+)}(k_{\textrm{SP}}|\mathbf{R}|)}{2}\hat{\mathbf{R}}\otimes\hat{\mathbf{R}}
\nonumber\\+\frac{H_1^{(+)}(k_{\textrm{SP}}|\mathbf{R}|)}{R}\hat{\bold{\phi}}\otimes\hat{\bold{\phi}}]
+ik_0k_{\textrm{SP}}H_1^{(+)}(k_{\textrm{SP}}|\mathbf{R}|)\hat{\mathbf{z}}\otimes\hat{\mathbf{R}}\}
\end{eqnarray} with
$\hat{\bold{\phi}}=\hat{\mathbf{z}}\times\hat{\mathbf{R}}$, and  $\delta_0=-\delta_1=1$.\\
\indent In the context of this work it is relevant to consider the
emission associated with a magnetic dipole since it is well known
that an aperture NSOM tip radiates like a coherent superposition of
in-plane electric and magnetic dipoles
$\mathbf{p}_{||,\textrm{aperture}}$ and
$\mathbf{m}_{||,\textrm{aperture}}=2\mathbf{p}_{||,\textrm{aperture}}\times\hat{\mathbf{z}}$,
with the electric dipole oriented along the electromagnetic mode
polarization at the tip apex~[19,23-26]. The SP Debye potential
associated with a radiating point-like magnetic dipole can be easily
obtained from Eq.~2. First, we slightly generalize Eq.~2 to describe
a current distribution
$\mathbf{J}(\mathbf{r'},z')=-i\omega\mathbf{P}(\mathbf{r'},z')$ and
obtain: $\Phi_{\textrm{SP};j}(\mathbf{r})=\int
d^{2}\mathbf{r'}dz'\eta_\bot(z')[H_0^{(+)}(k_{\textrm{SP}}|\mathbf{R}|)\mathbf{\hat{z}}\cdot\frac{\mathbf{J}(\mathbf{r'},z')}{-i\omega}
+i\frac{k_0}{k_{\textrm{SP}}}H_1^{(+)}(k_{\textrm{SP}}|\mathbf{R}|)\mathbf{\hat{R}}\cdot\frac{\mathbf{J}(\mathbf{r'},z')}{-i\omega}]$,
with $\mathbf{R}=\mathbf{r}-\mathbf{r}'$, and where the explicit
dependence of $\eta_\bot(z')$ over $z'$ is taken into account.
Second, we consider a magnetic loop, or equivalently, a current
distribution
$\mathbf{J}(\mathbf{r'},z')=c\nabla'\times\mathbf{M}(\mathbf{r'},z')$,
where $\mathbf{M}$ is the magnetic dipole distribution. After
integration by parts and use of the point-like dipole limit:
$\mathbf{M}(\mathbf{r'},z')=\mathbf{m}\delta^2(\mathbf{r'}-\mathbf{r}_0)\delta(z'-h)$
we obtain
\begin{eqnarray}
\Phi_{\textrm{SP}}(\mathbf{r})=-i\frac{k}{k_{\textrm{SP}}}\eta_{\bot}
H_1^{(+)}(k_{\textrm{SP}}|\mathbf{R}|)\mathbf{\hat{\phi}}\cdot\mathbf{m}_{||}.
\end{eqnarray} \\
\indent A few points are here remarkable. First, we find that for a
vertical magnetic dipole the SP field vanishes, a very fact that
qualitatively agrees with the symmetry of the magnetic field
$B_z\neq0$ associated with a vertical magnetic dipole, which cannot
couple to transverse magnetic waves, i.e., SPs for which $B_z=0$.
Second, and this is very important in the NSOM context, Eq.~4 is
equivalent to Eq.~2 if we define an effective electric dipole
$\mathbf{p}_{||,\textrm{eff.}}=\frac{k}{k_0}\hat{\mathbf{z}}\times\mathbf{m}_{||}$
since
$-\mathbf{\hat{\phi}}\cdot\mathbf{m}_{||}=\frac{k_0}{k}\mathbf{\hat{R}}\cdot\mathbf{p}_{||,\textrm{eff.}}$.
In other words it is not possible to distinguish the SP field
created by an in-plane magnetic dipole $\mathbf{m}_{||}$ from the
one created by an in-plane electric dipole
$\mathbf{p}_{||,\textrm{eff.}}$ obtained by rotating
$\mathbf{m}_{||}$ by $\pi/2$ around the $z$ axis. We think that this
difficulty could be of particular importance in the context of
measurements aiming at determining the dipole orientation of, e.g.,
a single nanohole drilled through a metal film~\cite{Rotenberg}.
Finally, due to the coefficient $\frac{k}{k_0}$ in the definition of
$\mathbf{p}_{||,\textrm{eff.}}$ we see that in general it is much
easier to launch a SP field with an in-plane magnetic dipole than
with an electric dipole of the same amplitude. For example at
$\lambda=647 nm$ using the previous value for
$\varepsilon_1(k)$~\cite{Johnson} we get on a semi infinite gold
film $(\frac{k}{k_0})^2\simeq (3.56)^2\simeq 12.7$. This is also in
qualitative agreement with the image dipole picture in which the
in-plane magnetic dipole and its image enforce each other while they
tend to compensate in the electric dipole case. Now, for the NSOM
tip we have a magnetic and an electric dipole orthogonal to each
other~[19,23-26]. Therefore, the total SP field created by such a
tip is equivalent to the field generated by an electric dipole
$\mathbf{p}_{||,\textrm{total}}=\mathbf{p}_{||,\textrm{aperture}}+\mathbf{p}_{||,\textrm{eff.}}=(1+2\frac{k}{k_0})\mathbf{p}_{||,\textrm{aperture}}$.
This agrees with experimental measurements of the SP
radiation profile for such a source~[12,14-17,20,21].\\
\indent To conclude this section we illustrate all discussed
features in Figs.~1 (a) and 1(b)  where we computed the SP intensity
(here $(\textrm{Re}[E_z])^2)$ along the air-metal interface for
respectively an horizontal electric dipole, an equivalent magnetic
dipole (as given by Eq.~4) and a vertical dipole. The selection
rules for exciting SPs are reminded in Fig.~1 (c) together with the
image dipole picture. Finally the SP field associated with an NSOM
tip is shown in Fig.~1 (d) together with the aperture dipole
directions.
\section{Plasmonic interferometry with a STM tip source }
\indent As a direct application we now analyze in more detail the
experimental configuration of ref.~\cite{sample}, where a STM tip
was used to excite SP waves on a 200 nm thick gold film. This film
was milled
\begin{figure}[h]
\begin{center}
\includegraphics[width=8.5cm]{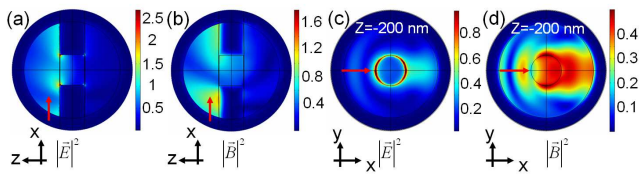}
\caption{Finite element simulation of the electromagnetic field
diffracted by a single cylindrical subwavelength aperture in a thick
gold film. The incident light mode is a pure SP wave with in-plane
wavevector aligned with the x axis (red arrow). (a) and (b) are
intensity maps in the x-z plane. (c) and (d) correspond to intensity
maps in the x-y plane ($z=-200$ nm). (a) and (c) are maps of the
total electric intensity $|\mathbf{E}(x,y,z)|^2$, whereas (b) and
(d) correspond to the magnetic intensity $|\mathbf{B}(x,y,z)|^2$. }
\label{figure5}
\end{center}
\end{figure}with two 250 nm-diameter holes separated by a distance of $d_{12}=2\mu$m.
In this experiment these holes acted as two coherent nano-sources of
light which subsequently interfered in the back-focal plane of a
high NA microscope objective. \\
\indent Let us first consider an isolated hole. Diffraction by
single hole in a metal film is a difficult theoretical problem with
a long history [50-52],  which  was recently renewed in the context
of plasmonics~\cite{NikitinPRL2010,Rotenberg,Yi,Bonod}. In order to
visualize the diffracted electromagnetic field in the vicinity of
the hole when a SP wave impinges on it, we first simulated the
problem by using a finite element analysis software [Comsol
Multiphysics] at the optical wavelength $\lambda=647 $ nm for the
material and geometry parameters of the STM
experiment~\cite{sample}. The results are shown in Fig.~2 for both
the total electric and magnetic field intensity $|\mathbf{E}|^2$ and
$|\mathbf{B}|^2$. The incident SP wave used in this simulation is a
rigorous mode of the air-metal-glass multilayer medium. As explained
in ref.~\cite{Genet,emrs} we consider here a leaky mode mainly
confined along the air-metal interface (i.e. $z=0$), which is
growing exponentially in the substratum and has an in-plane
wavevector $\textbf{k}_{\textrm{SP}}$ aligned along the +$x$
direction (direction indicated by a red arrow in Fig.~2). From the
images in Figs. 2(a) and 2(b) we see that the transmitted field is
strongly damped through the hole. This is reminiscent of previous
studies of transmission by holes, where only the fundamental mode
(i.e. in general the transverse electric mode
$\textrm{TE}_{11}$~\cite{Yi}) is excited. Here however, the incoming
SP is a pure TM wave and therefore the coupling with TM modes cannot
be neglected. Comparing the magnetic and electric intensity images
we see that $|\mathbf{E}|^2\sim |\mathbf{B}|^2$, in opposition with
results obtained with an incident propagating light wave instead of
a SP mode (compare for example with Fig. 1 of ref.~\cite{Kim5}).
This will naturally imply a different expansion into multipoles as
for the usual Bethe diffraction formula~\cite{Yi,Bethe,Roberts} (see
for example \cite{Rotenberg}, in which the vertical component
electric dipole $p_z$ plays a fundamental role). Finally, the
intensity profiles in the exit aperture plane (i.e. $z=-200$ nm)
reveal symmetries along the privileged direction defined by the SP
wavevector but no clear dominance of the magnetic field over the
electric field (see Figs.~2(c) and 2(d)) despite the fact that the
electric field is strongly confined near the
rim of the hole. This again stresses the fact that with SPs as excitation of single holes, both the magnetic and electric components play an important role.\\
\indent  It is beyond the aim of the present work to give a full
discussion of the SP diffraction problem by hole(s). Here, we
instead emphasize simple arguments adapted to our specific
problem. We will use a semi-analytical approach based on the Green Dyadic tensor formalism.\\
\indent For this purpose we remind that the electric displacement
field $\mathbf{D}(M)$ at point $M$ in the region surrounding the
hole is given by a Lippman Schwinger integral
$\mathbf{D}(M)=\mathbf{D}_{\textrm{SP}}(M)+\int_{V}d^{3}x'\bar{\textbf{G}}_{\textrm{film}}(M,M')(\varepsilon_0-\varepsilon_1)\mathbf{E}(M')$
where $\bar{\textbf{G}}_{\textrm{film}}(M,M')$ is the total dyadic
Green tensor corresponding to the film without hole, the integration
volume $V$ corresponds to the cylindrical region occupied by the
hole (filled with air) and $\mathbf{D}_{\textrm{SP}}(M)$ is the
incident SP field propagating along the interface $z=0$ and existing
without the hole~\cite{NikitinPRL2010,Yi}. In the transmitted region
$z<-d$, $\mathbf{D}_{\textrm{SP}}(M)\approx 0$ and only the volume
integral survives. Now, to first-order (Born) approximation, we can
write in the transmitted region $\mathbf{D}(M)\simeq
\int_{V}d^{3}x'\bar{\textbf{G}}_{\textrm{film}}(M,M')(\varepsilon_0-\varepsilon_1)\mathbf{E}_{\textrm{SP}}(M')$,
where $\mathbf{E}_{\textrm{SP}}$ is the incident unperturbed SP
field. However, since the SP field strongly decays in the metal
(since $1/(2\textrm{Imag}k_1)\simeq 10 nm$) the volume integral
evolves into a surface integral over the aperture area $S$:
$\mathbf{D}(M)\simeq\frac{i}{k_1}
\int_{S}d^{2}\mathbf{r}'\bar{\textbf{G}}_{\textrm{film}}(M,\mathbf{r}',z'=0^-)(\varepsilon_0-\varepsilon_1)\mathbf{E}_{\textrm{SP}}(\mathbf{r}',z'=0^-)$,
where the coefficient $\frac{i}{k_1}$ arises from an integration of
the SP exponential decay
$\int^0_{-h}dzf_1(z')\simeq\int^0_{-\infty}dzf_1(z')=i/k_1$.
Importantly, contrary to what occurs in air, the SP field in the
metal is dominated by its tangential components since
$|E_z|/|E_t|\simeq |k_{\textrm{SP}}/k_1|\simeq 0.27$ for the same
material parameter as before. Therefore, for small radius $a$ the
hole acts essentially as an in-plane dipole
$\mathbf{p}_{\textrm{hole}}\simeq\frac{i(\varepsilon_0-\varepsilon_1)}{k_1}
\int_{S}d^{2}x'\mathbf{E}_{\textrm{SP}}(\mathbf{r}',z'=0^-)$ leaking
through the metal film and located on the interface $z=0$. We point
out that this first-order approximation could lead to some problems
at large diffraction angle where it is known that leaky photons can
couple to SPs at the specific angle
$\theta_{\textrm{LR}}=\arcsin{(n_{\textrm{SP}}/n)}$~\cite{emrs,Genet}.
However, we will neglect this point here and will consider that for
horizontal dipoles, the dominant contribution is associated with the
radiative part occurring below the total internal reflection angle
$\theta_{c}=\arcsin{(1/n)}$.\\
In the next stage, to describe the diffracted field one must also
take into account the hole size and retardation effects. For this
purpose, consider now a point-like dipole $\mathbf{p}$ located in
the hole and radiating light in free space in the direction of the
microscope objective. The electric field measured at a distance
$S=\textrm{OM}\gg\lambda$ of the plane $z=0$ takes the asymptotic
form
$\mathbf{E}(t)\simeq\textrm{const.}\frac{p(t-nS/c)}{S}[\mathbf{\hat{p}}-(\mathbf{\hat{p}}\cdot\mathbf{\hat{S}})\mathbf{\hat{S}}]$,
where $\mathbf{\hat{S}}=\bold{OM}/|\bold{OM}|$ is the unit vector
directed from the hole center O to the observation point M and $n$
is the optical index of the substrate (this formula is justified
since the asymptotic part of $\bar{\textbf{G}}_{\textrm{film}}$ for
observation points not too far from the optical axis $-z$ approaches
the dyadic for the substratum of permittivity $n^2$). From the
previous discussion it should now be clear that the aperture as a
whole acts as a coherent integral distribution of such dipoles
excited by the incident (in-plane) SP field
$\mathbf{E}_{\textrm{SP}}$ impinging on the hole. The total field at
the observation point reads consequently
$\mathbf{E}=\textrm{const.}\frac{e^{iknS}}{S}\mathbf{Q}[\mathbf{k}_{||}]$,
where $\mathbf{Q}[\mathbf{k}_{||}]$ is the structure factor defined
by
\begin{eqnarray}
\mathbf{Q}[\mathbf{k}_{||}]=\int_{S}d^{2}\mathbf{r}'e^{-i\mathbf{k}_{||}\cdot\mathbf{r}'}[\mathbf{E}_{\textrm{SP}}(\mathbf{r}')-(\mathbf{E}_{\textrm{SP}}(\mathbf{r}')\cdot\mathbf{\hat{S}})\mathbf{\hat{S}}]
\end{eqnarray} calculated for the special in-plane wavevector $\mathbf{k}_{||}=kn[(\mathbf{\hat{S}}\cdot\mathbf{\hat{x}})\mathbf{\hat{x}}+(\mathbf{\hat{S}}\cdot\mathbf{\hat{y}})\mathbf{\hat{y}}]$.
Suppose for example an incident SP wave
$\mathbf{E}_{\textrm{SP},||}(\mathbf{r}')=\mathbf{\hat{u}}e^{ik_{\textrm{SP}}\mathbf{\hat{u}}\cdot\mathbf{r}'}$
(i.e., considering only in-plane components of the SP field), and a
cylindrical hole centered on the in-plane vector
$\mathbf{r}_0=x_0\mathbf{\hat{x}}+y_0\mathbf{\hat{y}}$ we deduce
\begin{eqnarray}
\mathbf{Q}[\mathbf{k}_{||};\mathbf{\hat{u}},\mathbf{r}_0]=[\mathbf{\hat{u}}-(\mathbf{\hat{u}}\cdot\mathbf{\hat{S}})\cdot\mathbf{\hat{S}}]2\pi
a^2e^{-i\mathbf{k}_{||}\cdot\mathbf{r}_0}\frac{J_1(va)}{va}
\end{eqnarray} with $v=|\mathbf{k}_{||}-k_{\textrm{SP}}\mathbf{\hat{u}}|$ (here we neglected the effect of the SP propagation length $L_{\textrm{SP}}\simeq 10$ $\mu$m in the integration since the hole diameter $2a$ is supposed to be much smaller).\\
\indent Finally, a rigorous description of the diffracted field
requires to consider the electromagnetic field distortion induced by
the propagation through the objective. This effect is expected to be
small for paraxial light rays propagating close to the optical axis
(i.e. in the $-z$ direction) but cannot be neglected in general with
high NA objectives. For this purpose, we apply the general  Debye
procedure described for example in refs.~\cite{Genet,Sheppard} that
consists in introducing a projection condition from the spherical
wave front going away from the sample plane onto a planar wave front
transmitted by the infinity corrected aplanatic objective. For the
source considered here the field on this spherical wave front just
before the transformation is collinear to
$\mathbf{\hat{u}}-(\mathbf{\hat{u}}\cdot\mathbf{\hat{S}})\cdot\mathbf{\hat{S}}$
, which can also be written as
$(\mathbf{\hat{u}}\cdot\mathbf{\hat{\theta}})\mathbf{\hat{\theta}}+(\mathbf{\hat{u}}\cdot\mathbf{\hat{\phi}})\mathbf{\hat{\phi}}$
using the spherical coordinate basis $[S,\theta,\phi]$. This vector
is clearly tangential to the sphere, as it should be for far-field
radiation. The projection on the planar wave front transforms this
vector into
$\frac{[((\mathbf{\hat{u}}\cdot\mathbf{\hat{\rho}})\cos{\theta}-(\mathbf{\hat{u}}\cdot\mathbf{\hat{z}})\sin{\theta})\mathbf{\hat{\rho}}+(\mathbf{\hat{u}}\cdot\mathbf{\hat{\phi}})\mathbf{\hat{\phi}}]}{\sqrt{(\cos{(\theta)})}}$
and the square root $\sqrt{(\cos{(\theta)})}$ is introduced to
ensure energy conservation during propagation~\cite{Sheppard}.
Therefore, using this relation, Eq.~6 must now be replaced by
\begin{eqnarray}
\mathbf{Q'}[\mathbf{k}_{||};\mathbf{\hat{u}},\mathbf{r}_0]=[((\mathbf{\hat{u}}\cdot\mathbf{\hat{\rho}})\cos{\theta}
-(\mathbf{\hat{u}}\cdot\mathbf{\hat{z}})\sin{\theta})\mathbf{\hat{\rho}}\nonumber\\+(\mathbf{\hat{u}}\cdot\mathbf{\hat{\phi}})\mathbf{\hat{\phi}}]\frac{2\pi
a^2e^{-i\mathbf{k}_{||}\cdot\mathbf{r}_0}}{\sqrt{(\cos{(\theta)})}}\frac{J_1(va)}{va}
\end{eqnarray} where $\mathbf{k}_{||}=kn\sin{\theta}\mathbf{\hat{\rho}}$. The electric field in the back-focal plane of the objective (Fourier plane) reads consequently
$\mathbf{E}=\textrm{const'.}\frac{e^{iknf}}{f}\mathbf{Q}'[\mathbf{k}_{||};\mathbf{\hat{u}},\mathbf{r}_0]$
where $f$ is the radius of the spherical wave front (reference
sphere), which corresponds
to the focal length of the objective.\\
\indent We now go back to the STM experiment considered in
ref.~\cite{sample} and suppose two holes located at
$\mathbf{r}_1=x_1\mathbf{\hat{x}}+y_1\mathbf{\hat{y}}$  and
$\mathbf{r}_2=x_2\mathbf{\hat{x}}+y_2\mathbf{\hat{y}}$ on the metal
film while the STM tip is located at
$\mathbf{r}_0=x_0\mathbf{\hat{x}}+y_0\mathbf{\hat{y}}$. SPs launched
from the tip to both apertures take the asymptotic form
$\textbf{E}_{\textrm{SP},1}\propto
\frac{\mathbf{r}_1-\mathbf{r}_0}{|\mathbf{r}_1-\mathbf{r}_0|}\frac{e^{ik_{\textrm{SP}}|\mathbf{r}_1-\mathbf{r}_0|}}{\sqrt{|\mathbf{r}_1-\mathbf{r}_0|}}$,
and $\textbf{E}_{\textrm{SP},2}\propto
\frac{\mathbf{r}_2-\mathbf{r}_0}{|\mathbf{r}_2-\mathbf{r}_0|}\frac{e^{ik_{\textrm{SP}}|\mathbf{r}_2-\mathbf{r}_0|}}{\sqrt{|\mathbf{r}_2-\mathbf{r}_0|}}$
for holes 1 and 2  respectively (the vertical component of the SP
fields are not considered here in agreement with our previous
discussion).
\begin{figure}[h]
\begin{center}
\includegraphics[width=8.5cm]{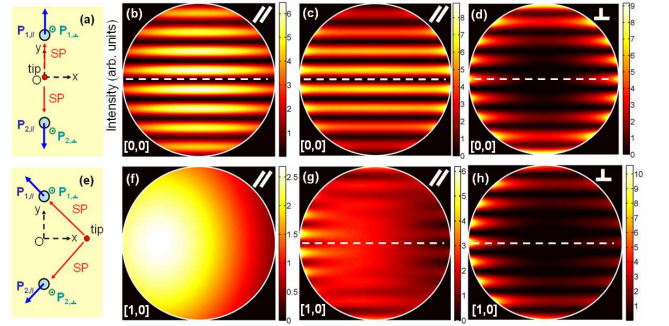}
\caption{Theoretical modeling of the SP excitation of two
subwavelength apertures by a STM tip. The first line (b-d) gives
intensity maps of the diffracted  intensity detected in the
back-focal (Fourier) plane  of the high NA objective (here $NA=1.4$)
corresponding to the configuration sketched in (a) when the STM  tip
is located midway between the holes. The second line (f-h)
corresponds to the configuration shown in (e). The blue,
respectively green, arrows in (a) and (e) represent the direction of
horizontal, respectively, vertical electric dipoles associated with
the holes. Images (b), (c), (f) and (g) correspond respectively to
the horizontal and vertical dipole case. The white dashed lines
along the $k_x$ axis shows the minimum or maximum fringe intensity.
} \label{figure5}
\end{center}
\end{figure}
The total electric field in the back-focal plane of the objective is
thus proportional to
\begin{eqnarray}
\textbf{E}(\mathbf{k}_{||})\propto
\frac{e^{ik_{\textrm{SP}}|\mathbf{r}_1-\mathbf{r}_0|}}{\sqrt{|\mathbf{r}_1-\mathbf{r}_0|}}\mathbf{Q'}[\mathbf{k}_{||};\frac{\mathbf{r}_1-\mathbf{r}_0}{|\mathbf{r}_1-\mathbf{r}_0|},\mathbf{r}_1]
\nonumber\\+\frac{e^{ik_{\textrm{SP}}|\mathbf{r}_2-\mathbf{r}_0|}}{\sqrt{|\mathbf{r}_2-\mathbf{r}_0|}}\mathbf{Q'}[\mathbf{k}_{||};\frac{\mathbf{r}_2-\mathbf{r}_0}{|\mathbf{r}_2-\mathbf{r}_0|},\mathbf{r}_2].
\end{eqnarray}
The intensity recorded in the back-focal plane is finally obtained
as $I(\mathbf{k}_{||})\propto|\mathbf{E}|^2$. This takes a simple
analytical form for small $\pi-\theta$, i.e., near the optical axis
$-z$, and in the limit of small radius $a$ for which the Airy
function $\frac{2J_1(va)}{va}\simeq 1$. Indeed, in that case, we get
\begin{eqnarray}
I(\mathbf{k})\propto\frac{e^{-R_1/L_{\textrm{SP}}}}{R_1}+
\frac{e^{-R_2/L_{\textrm{SP}}}}{R_2}\nonumber\\+2\frac{\mathbf{\hat{R}}_1\cdot\mathbf{\hat{R}}_2}{\sqrt{R_1R_2}}e^{-\frac{R_1+R_2}{2L_{\textrm{SP}}}}\cos{[\mathbf{k}_{||}\cdot\mathbf{d}_{12}+kn_{\textrm{SP}}(R_2-R_1)]}
\end{eqnarray} with $\mathbf{R}_1=\mathbf{r}_1-\mathbf{r}_0$, $\mathbf{R}_2=\mathbf{r}_2-\mathbf{r}_0$ and $\mathbf{d}_{12}=\mathbf{r}_1-\mathbf{r}_2$.\\
\indent As an illustration, we now consider the two cases sketched in Fig.~3(a) and 3(e) where:\\
i) $\mathbf{R}_1=-\mathbf{R}_2$, i.e., the tip is located along the line joining the holes at mid-way from each (Fig.~3(a)).\\
ii) The vectors $\mathbf{R}_1$ and $\mathbf{R}_2$ are orthogonal and equal in norms (Fig.~3 (e)).\\
\indent In configuration i) the fringe visibility predicted by Eq.~9
is maximal in norm: $V=-1$ while the phase shift $\Delta \phi
=kn_{\textrm{SP}}(R_2-R_1)$ vanishes. Oppositely, in configuration
ii) Eq.~9 predicts no interference since the fringes visibility
vanishes. This result is reproduced by the simulation taking into
account the full expression of
$\mathbf{Q}[\mathbf{k}_{||};\mathbf{\hat{u}}]$ (Eq.~6) as shown in
Figs.~3(b) and 3(f). This corresponds to the intensity that would be
measured before the transmission by the objective, i.e., for
observation points confined on the reference sphere of radius $f$.
This could be experimentally recorded using a
goniometer~\cite{Karrai2,Yi}. In particular, while there is no
interference for case ii) (see Fig.~3(f)) the value $V=-1$ in case
i) imposes an interference fringe minimum along the axis $k_y=0$
(the white dashed line in Fig.~3(b)), in agreement with the
simplified model discussed above. The same features are also
observed in the back-focal objective plane if we consider
$\mathbf{Q'}[\mathbf{k}_{||};\mathbf{\hat{u}},\mathbf{r}]$ instead
of $\mathbf{Q}[\mathbf{k}_{||};\mathbf{\hat{u}},\mathbf{r}]$,~i.e.
Eqs.~7,8 (see Figs.~3(c) and (g)).
The distortions observed in Fig.~3(g) compared to Fig.~3(f) are related to the mixing between the polarization induced by the microscope objective and there is now a small fringe minimum along the axis $k_y=0$ (white dashed line in Fig.~2(g)).\\
\indent For comparison, we also show the simulated images in the
Fourier space for the case ii) if the holes are supposed to react
like vertical dipoles instead of  horizontal dipoles (i.e.
Figs.~3(d) and 3(h)). In analogy with the discussion leading to
Eq.~6 this means that the hole is now driven by the vertical
component of the incoming SP:
$\mathbf{E}_{\textrm{SP},\bot}(\mathbf{r}')=\mathbf{\hat{z}}e^{ik_{\textrm{SP}}\mathbf{\hat{u}}\cdot\mathbf{r}'}$.
This situation is automatically predicted using Eqs.~6,7 and
replacing $\mathbf{\hat{u}}_1$, $\mathbf{\hat{u}}_2$ by the
direction $\mathbf{\hat{z}}$ everywhere but in the definition of
$v$. In particular using the same approximations as in Eq.~9 we find
that the intensity should vanish in the paraxial regime since to
zero order $\mathbf{\hat{z}}\sim\mathbf{\hat{S}}$ and
$\mathbf{\hat{z}}-(\mathbf{\hat{z}}\cdot\mathbf{\hat{S}})\cdot\mathbf{\hat{S}}\simeq
0$. A more precise calculation imposing only
$\frac{2J_1(va)}{va}\simeq 1$ however naturally leads to
\begin{eqnarray}
I(\mathbf{k})\propto\frac{(\sin{\theta})^2}{\cos{\theta}}\{\frac{e^{-R_1/L_{\textrm{SP}}}}{R_1}+
\frac{e^{-R_2/L_{\textrm{SP}}}}{R_2}\nonumber\\+2\frac{1}{\sqrt{R_1R_2}}e^{-\frac{R_1+R_2}{2L_{\textrm{SP}}}}\cos{[\mathbf{k}_{||}\cdot\mathbf{d}_{12}+kn_{\textrm{SP}}(R_2-R_1)]}\}
\end{eqnarray} in the back-focal plane of the objective.
This is confirmed by the complete simulation using an adapted
structure factor, i.e., replacing Eq.~7 in Eq.~8 by
$\mathbf{Q''}[\mathbf{k}_{||};\mathbf{\hat{z}},\mathbf{\hat{u}},\mathbf{r}]=\frac{[-\sin{\theta}\mathbf{\hat{\rho}}]}{\sqrt{(\cos{(\theta)})}}2\pi
a^2e^{-i\mathbf{k}_{||}\cdot\mathbf{r}_0}\frac{J_1(va)}{va}$ (as
shown in Figs.~3d and 3h). Importantly, we find in both cases i) and
ii) a residual visibility along the axis $k_y=0$, visible only at
large $\theta$, due to the optical distortion through the microscope
(i.e., the $\sin{\theta}^2$ term).\\
Therefore, we conclude this
section by suggesting that the STM SP point source used in
refs.~\cite{STM1,STM2,sample} constitutes an ideal system for
probing the dipole orientations associated with diffracting holes in
a thick metal film.
\section{Plasmonic interferometry with a NSOM tip source}
\indent In the case of a NSOM SP source the previous configuration
with two holes is not ideal since the in-plane dipoles, either
electric or magnetic, associated with the tip aperture introduce an
additional degree of freedom which can affect the fringe visibility
in the back-focal plane of the microscope. Here, rather than a pair
of holes, we consider a circular slit of diameter $D=3,4$ or $5$
$\mu$m and width $w=250$ nm milled in a 200 nm thick gold film: see
Figs.~4(a), 4(b), and 4(c), respectively. This circular symmetry
restores the symmetry lost with the replacement of the vertical tip
dipole by two in-plane orthogonal dipoles. In order to give a
detailed description of what is going on when the NSOM tip is
located at $\mathbf{r}_0=x_0\mathbf{\hat{x}}+y_0\mathbf{\hat{y}}$
inside this circular corral we first parameterize any point along
the aperture rim of radius $D/2$ by
$\mathbf{r}(\beta)=(D/2)[\cos{\beta}\mathbf{\hat{x}}+\sin{\beta}\mathbf{\hat{y}}]$
with $\beta\in [0,2\pi[$. As shown by Eqs.~2 or 3, the SP field
generated at such a running point by the equivalent electric dipole
$\mathbf{p}_{||,\textrm{total}}$ associated with the tip takes the
asymptotic form
$\mathbf{E}_{\textrm{SP}}\propto\frac{e^{ik_{\textrm{SP}}|\mathbf{R}(\beta)|}}{\sqrt{|\mathbf{R}(\beta)|}}(\mathbf{p}_{||,\textrm{total}}\cdot\mathbf{\hat{R}}(\beta))\mathbf{\hat{R}}(\beta))$
with $\mathbf{R}(\beta)=\mathbf{r}(\beta)-\mathbf{r}_0$ (the
different vectors involved in this analysis are represented in
Fig.~4(c) for clarity).\\
Now, in contrast with a single hole, a
single slit reacts anisotropically to an incoming SP wave. Indeed,
it has been shown experimentally that a linear slit acts as a
polarizer transmitting or scattering light only if the incoming
in-plane SP wavevector $\mathbf{k}_{\textrm{SP}}$ is normal to the
slit (see e.g.~\cite{BaudrionPRB,Degiron}). This has also been
confirmed with circular slits to tailor specific polarization
convertors~\cite{Lombard,Drezetpola}. Here, it implies that each
running point $\mathbf{r}(\beta)$ acts a dipole transmitting light
in the direction $\mathbf{\hat{r}}(\beta)$ with an amplitude
proportional to the scalar product
$\mathbf{\hat{R}}(\beta)\cdot\mathbf{\hat{r}}(\beta)$ in agreement
with Malus's law. Therefore, by linear superposition of all these
point-like sources, i.e. after integration over $\beta$, we get for
the field recorded in the back-focal plane:
\begin{eqnarray}
\textbf{E}(\mathbf{k}_{||})= \textrm{const.}
\oint\frac{e^{ik_{\textrm{SP}}|\mathbf{R}(\beta)|}}{\sqrt{|\mathbf{R}(\beta)|}}(\mathbf{p}_{||,\textrm{total}}\cdot\mathbf{\hat{R}}(\beta))\nonumber\\
\mathbf{Q'}[\mathbf{k}_{||};(\mathbf{\hat{R}}(\beta)\cdot\mathbf{\hat{r}}(\beta))\mathbf{\hat{r}}(\beta),\mathbf{r}(\beta)]d\beta
\end{eqnarray}
in complete analogy with Eq.~8.\\
\indent We \begin{figure}[h]
\begin{center}
\includegraphics[width=8.5cm]{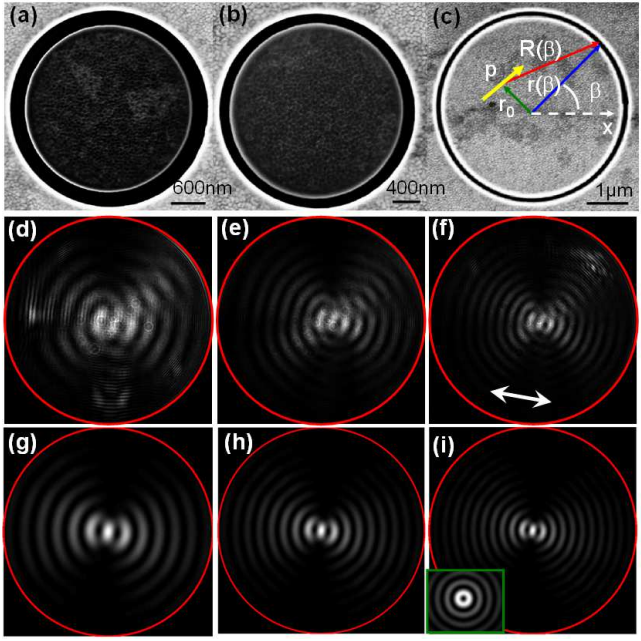}
\caption{SP excitation of a ring like aperture by a NSOM tip.  (a-c)
show FIB images of the sample considered with ring diameters 3, 4
and 5 $\mu$m. The directions of the vectors used in the mathematical
model are shown in (c). (d-f) are intensity maps in the Fourier
plane  and  correspond to samples (a-c) when the tip is located
nearly at the ring cavity center. (g-i) are the associated
calculated images for the equivalent horizontal electric NSOM dipole
indicated by a double white arrow. The inset in (i) shows what would
be observed with a STM tip (images are at the same scale).}
\label{figure5}
\end{center}
\end{figure}consider first the case where the NSOM tip is at the center of the cavity. In that situation the recorded electric field can be approximately evaluated if
we write
$\mathbf{\hat{u}}-(\mathbf{\hat{u}}\cdot\mathbf{\hat{S}})\cdot\mathbf{\hat{S}}\simeq
\mathbf{\hat{u}}$ (paraxial ray approximation) in Eq.~6 or 7. We
obtain
\begin{eqnarray}
\textbf{E}(\mathbf{k}_{||})=\textrm{const.}\frac{\pi\sqrt{2}e^{ik_{\textrm{SP}}D/2}}{\sqrt{D}}\{[J_0(k_{||}D/2)\nonumber\\-J_2(k_{||}D/2)((\hat{\mathbf{k}}_{||}\cdot\mathbf{\hat{p}}_{||})^2
-(\hat{\mathbf{k}}_{||}\times\mathbf{\hat{p}}_{||})^2)]\hat{\mathbf{k}}_{||}\nonumber\\-2J_2(k_{||}D/2)(\hat{\mathbf{k}}_{||}\cdot\mathbf{\hat{p}}_{||})(\hat{\mathbf{k}}_{||}\cdot(\hat{\mathbf{z}}\times\mathbf{\hat{p}}_{||}))\hat{\mathbf{z}}\times\mathbf{\hat{p}}_{||}\}\nonumber\\
\end{eqnarray} i.e., \begin{eqnarray}
\textbf{E}(\mathbf{k}_{||})\simeq\textrm{const.}\frac{\pi\sqrt{2}e^{ik_{\textrm{SP}}D/2}}{\sqrt{D}}\{[J_0(k_{||}D/2)\nonumber\\-J_2(k_{||}D/2)\cos{(2\phi)}]\hat{\mathbf{x}}-J_2(k_{||}D/2)\sin{(2\phi)}\hat{\mathbf{y}}\}
\end{eqnarray} where $\mathbf{k}_{||}=k_{||}(\cos{\phi}\hat{\mathbf{x}}+\sin{\phi}\hat{\mathbf{y}})$.
In going from Eq.~12 to 13 we imposed $\mathbf{\hat{p}}_{||}$ to
simplify the mathematical expression (i.e. the equivalent NSOM
electric dipole is aligned with the x axis). For comparison, if we
consider a STM tip (i.e. with a dipole source aligned with the $z$
axis) instead of a NSOM source and do the calculation at the same
degree of precision we obtain :
\begin{eqnarray}
\textbf{E}(\mathbf{k}_{||})\simeq
\textrm{const.}\frac{-2i\pi\sqrt{2}e^{ik_{\textrm{SP}}D/2}}{\sqrt{D}}J_1(k_{||}D/2)\hat{\mathbf{k}}_{||}.
\end{eqnarray}
What is noticeable in comparing Eq.~13 and 14 is the presence of
Bessel functions which are reminiscent of the cylindrical waves
already studied in the context of plasmonics for optical states of
polarization
conversion~\cite{Zhang,Odom,Hasman,Lombard,Drezetpola,Yuribis}. In
particular, the Bessel function $J_1(k_{||}D/2)$ appearing in Eq.~13
involves a vortex of topological charge
$\pm1$~\cite{Lombard,Yuribis} with an intensity minimum for
$k_{||}=0$,~a fact which is in complete agreement with the radial
symmetry of the STM tip field along the interface $z=0$. In
contrast, the contributions to the signal in the case of the NSOM
tip is split between a fundamental $J_0(k_{||}D/2)$ term and an
optical vortex of topological charge $\pm2$ due to the difference of
symmetry of the electric field at the tip apex. This results into a
maximum of intensity at $k_{||}=0$.\\
This prediction is
\begin{figure}[h]
\begin{center}
\includegraphics[width=8.5cm]{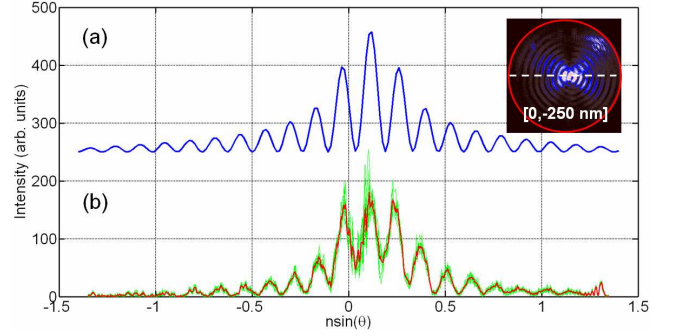}
\caption{Intensity crosscuts  along $k_x$ of the image shown in
Fig.~4 (f) [see also inset, where the white dashed line indicates
the crosscut direction while the actual tip coordinates $[x,y]$ are
indicated in brackets]. (a): Theoretical modeling deliberately
shifted in the intensity direction for clarity. (b): The green
curves shown  correspond to several crosscuts  and the red curve
gives their average. } \label{figure5}
\end{center}
\end{figure}confirmed experimentally as shown in Fig.~4 for the 3 different diameters $D$. The tip is approximately located
at the center of the ring cavity and the simulation are realized
with the full Eq.~11. The good agreement between experiments
(Figs.~4(d), 4(e), 4(c)) and our model (Figs.~4(g), 4(h), 4(i))
demonstrates the  validity of our semi-analytical description. We
point out that the vertical dipole STM tip would lead to completely
different images. As an example, the numerical calculation for the
STM tip placed at the center of the corral is shown in the inset of
Fig.~4 (i). It indeed reveals a vortex like structure with
topological charge $\pm1$ in clear disagreement with the experiment
(Fig.~4(f)). A better agreement is even obtained if we consider the
actual position of the tip in Fig.~4(i) which is slightly off the
center (by an amount of 250 nm) along the x axis. This is shown in
Fig.~5 where crosscuts were taken along, and parallel (i.e. very
close) to, the $k_x$ axis of the intensity mapped in Fig.~4(i).  The
average intensity obtained is in good agreement with the theoretical
predictions of our model. In this context it is also interesting to
calculate the scattered field  in the Fourier space with the
hypothesis that the circular slit responds as a distribution of
vertical dipoles instead of in-plane dipoles, whereas the NSOM tip
still behaves like an in-plane dipole. In analogy with the
discussion preceding Eq.~10, we get here:
\begin{eqnarray}
\textbf{E}(\mathbf{k}_{||})\simeq
\textrm{const.}\frac{2i\pi\sqrt{2}e^{ik_{\textrm{SP}}D/2}}{\sqrt{D}}\nonumber\\
J_1(k_{||}D/2)(\hat{\mathbf{k}}_{||}\cdot\hat{\mathbf{p}}_{||})\hat{\mathbf{k}}_{||}\frac{\sin{\theta}}{\sqrt{\cos{\theta}}}
\end{eqnarray}
If we drop the smooth $\sin{\theta}$ variation, this result is
intermediate between Eq.~14 and 13 since it combines a vortex of
topological charge $\pm1$ with a cos like pattern (i.e. the term
$(\hat{\mathbf{k}}_{||}\cdot\hat{\mathbf{p}}_{||})$). Such behavior
is clearly excluded by the data shown in Figs.~4 and 5.
\begin{figure}[h]
\begin{center}
\includegraphics[width=8.5cm]{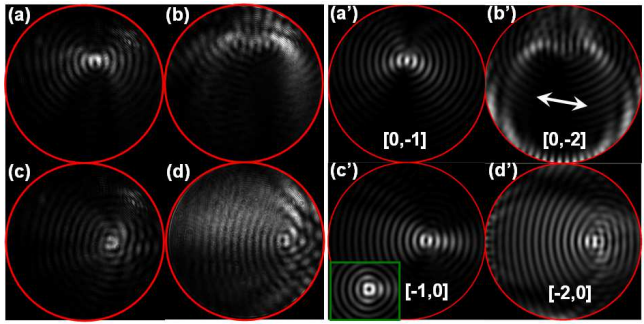}
\caption{SP excitation of two subwavelength apertures by a
ex-centered NSOM tip. (a-d) correspond to the experimental Fourier
space images while (a'-d') are the associated theoretical images.
The white arrow indicates the tip polarization and the brackets
indicate the $[x,y]$ tip coordinates.   } \label{figure5}
\end{center}
\end{figure}All this analysis therefore confirms
that in the considered case (Figs. 4  and 5) the  circular slit acts locally as a sort of Malus polarizer for the in-plane SP electric field.\\
\indent As a final analysis it is relevant to study further the
Fourier space images when the NSOM tip is not located at the center
of the circular cavity. This is shown in Fig.~6 for a $D=5$ $\mu$m
diameter cavity and for the tip displaced along the x or  y axis
respectively by an amount of approximately 1 or 2 $\mu$m (i.e.,
still inside the cavity). The data of Fig. 6(a-d) are compared in
Fig. 6(a'-d') with the theoretical model using  Eq.~11. Again, a
good agreement is recovered despite some differences at large angles
\begin{figure}[h]
\begin{center}
\includegraphics[width=8.5cm]{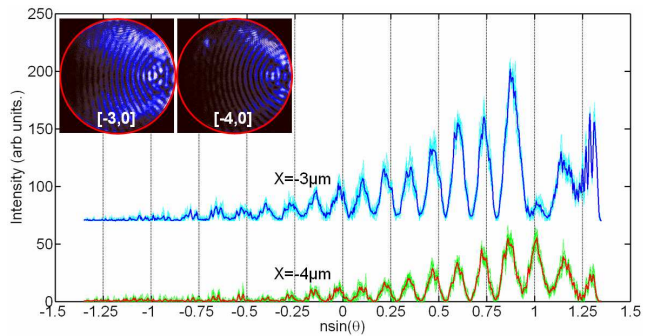}
\caption{Intensity crosscuts  along $k_x$ of the images shown in the
inset. The white dashed line indicates the crosscut direction and
the actual coordinate $[x,y]$ of the tip are indicated in brackets.
Crosscuts corresponding to $x=-3$ and $-4$ $\mu$m have be shifted
along the intensity axis for clarity.  } \label{figure5}
\end{center}
\end{figure}attributed to either optical misalignments or to some NSOM tip asymmetries. The most interesting feature is the observation of SP fringes of typical periodicity $\delta k=2\pi/D$ observed
in Fig. 6(d) (and recovered in the simulation  of Fig.~6 (d')) when
the tip is displaced along the x axis. These fringes are absent if
the tip is displaced along the y axis. This is explained by the fact
that the tip polarisation is mainly oriented along the x axis.
Therefore, SPs couple favorably to the slit parts intersecting the x
axis and practically do not couple along the y axis. The system
actually acts like a Young double slit or hole experiment in strong
analogy with what was obtained in refs.~\cite{Gan,Kuzmin} and
particularly ref.~\cite{sample}. However, in contrast with the
double hole experiment, here the light sources are spatially
extended and curved so that a simple analysis identical to what was
done in Section 3 and in ref.~\cite{sample} is not possible. Still,
we show in Fig.~7 Fourier space images obtained when the  NSOM tip
is now outside the ring cavity i.e. displaced by an amount of 3 of 4
$\mu$m along the x axis. Crosscuts made along the $k_x$ axis in the
Fourier space  show first that the fringe periodicity
$\delta(n\sin{\theta})=\lambda/D\simeq 0.13$ obtained in the
double-slit experiment is indeed recovered. Second, we observe that
the fringe minimum and maximum do not move when we go from the case
where the NSOM tip is located at $x=-3$ $\mu$m to the position
$x=-4$ $\mu$m. This is explained by the fact that the phase shift
between the two parts of the slit contributing to the Fourier space
image (i.e., those parts of the slit intersecting the x axis) is
given in magnitude by $kn_{\textrm{SP}}|R_1-R_2|$, where $R_1$ and
$R_2$ are respectively the distance between the tip and the part of
the slit located at $x\simeq D/2$ and $x\simeq -D/2$ (see Eq.~9).
Since $R_1-R_2=D$ when the tip is outside the cavity  this phase
shift is the same in the two cases considered and no fringe
displacement is expected. This is also in qualitative agreement with
the model which reproduces this main feature.
\section{Conclusion}
In this article we have developed a semi-analytical approach able to
analyze plasmonic interferometry using scanning near-field SP
sources of two kinds: STM and NSOM tips. Our method, mainly based on
a scalar Debye-Green formalism for electric and magnetic dipole
sources with proper consideration of the propagation through the
collection objective, was specifically adapted to describe recent
experiments using a STM source in a Young double hole experiment.
The model clearly reproduces the observed features~\cite{sample} and
predicts some interesting configuration which could be used in the
future to probe the aperture dipole directionality.  We also focused
our attention, both experimentally and theoretically, on plasmonic
interferometry on a circular slit achieved with a NSOM source. The
measurements were successfully compared to our model, thereby
proving the efficiency of our approach. We expect this work to have
important applications in the field of plasmonics where optical
vortices and manipulation of the light polarization are coupled to
near-field microscopy~[11,46-48].

\indent Acknowledgements: We thank  Jean-Fran\c{c}ois Motte, from
NANOFAB facility in Neel Institute for the optical tip manufacturing
and FIB milling of the circular slits used in this work. This work
was supported by Agence Nationale de la Recherche (ANR), France,
through the PLASTIPS, SINPHONIE and PLACORE
projects.\\

\end{document}